\documentclass{raa_twocolumn} 
\usepackage{pdflscape}
\usepackage{natbib}
\usepackage{amssymb,amsmath}
\usepackage{longtable}
\bibpunct{(}{)}{;}{a}{}{,}
\usepackage{graphicx}
\usepackage[a4paper=true,pagebackref=true]{hyperref}
\hypersetup{pdftitle = The title of my PDF, pdfauthor = My name, pdfsubject= The subject, pdfkeywords = keyword1 keyword2 keyword3}
\hypersetup{colorlinks = true, linkcolor = blue, anchorcolor = blue, citecolor = blue, filecolor = blue, pagecolor = blue, urlcolor = blue}
\footnotesize
\newdimen\digitwidth    
\setbox0=\hbox{\rm0}
\digitwidth=\wd0
\catcode`!=\active
\def!{\kern\digitwidth}
\normalsize

\usepackage{soul}

\begin{document}

\title{A new emission mode of PSR B1859+07}

\volnopage{ {\bf 2022} Vol.\ {\bf 21} No. {\bf 1}, A01}
\setcounter{page}{1}
\author{
    Tao~Wang\inst{1,2}, P.~F.~Wang\inst{1,2,3}, J.~L.~Han\inst{1,2,3}, Yi Yan\inst{1,2}, Ye-Zhao Yu\inst{4}, Feifei Kou\inst{5}  
   }
\institute{
National Astronomical Observatories, Chinese Academy of Sciences, 20A Datun Road, Chaoyang District, Beijing 100012, China;  {\it pfwang, hjl@nao.cas.cn}\\
\and
School of Astronomy and Space Science, University of Chinese Academy of Sciences, 19A Yuquan Road, Beijing 100049, China\\
\and     
The key Laboratory for Radio Astronomy and Technology, Chinese Academy of Sciences, Jia20 Datun Road, Beijing 100012, China; \\
\and
Qiannan Normal University for Nationalities, Duyun, 558000, China; \\
\and
Xinjiang Astronomical Observatories, Chinese Academy of Sciences, Urumqi, 830011, China
\vs \no
{\small Received 2021 XXX; accepted 2021 XXX}
}

\abstract{
Previous studies have identified two emission modes in PSR B1859+07: a normal mode that has three prominent components in the average profile, with the trailing one being the brightest, and an anomalous mode (i.e. the A mode)  where emissions seem to be shifted to an earlier phase. Within the normal mode, further analysis has revealed the presence of two sub-modes, i.e. the cW mode and cB mode, where the central component can appear either weak or bright. As for the anomalous mode, a new bright component emerges in the advanced phase while the bright trailing component in the normal mode disappears. New observations of PSR B1859+07 by using the Five-hundred-meter Aperture Spherical radio Telescope (FAST) have revealed the existence of a previously unknown emission mode, dubbed as the Af mode. In this mode, all emission components seen in the normal and anomalous modes are detected. Notably, the mean polarization profiles of both the A and Af modes exhibit an orthogonal polarization angle jump in the bright leading component. The polarization angles for the central component in the original normal mode follow two distinct orthogonal polarization modes in the A and Af modes respectively. The polarization angles for the trailing component show almost the same but a small systematic shift in the A and Af modes, roughly following the values for the cW and cB modes. Those polarization features of this newly detected emission mode imply that the anomalous mode A of PSR B1859+07 is not a result of ``phase shift" or ``swooshes" of normal components, but simply a result of the varying intensities of different profile components. Additionally, subpulse drifting has been detected in the leading component of the Af mode. 
}
\authorrunning{Tao Wang et al.}            
\titlerunning{A new emission mode of PSR B1859+07}  

\maketitle
%



\section{Introduction}

Pulsars are highly magnetized, rotating neutron stars. The pulses generated from each rotation exhibit varying morphology and polarization. Pulse sequences manifest nulling, mode changing, and subpulse drifting phenomena. 
\citet{rrw+2006} found an unusual phenomenon that PSRs B0919+06 and B1859+07 occasionally manifests as a shift of emission towards early rotation phases, which was termed as ``swooshes" by \citet{wor+2016} and \citet[e.g.][]{rps+2021} or the ``flare-state'' by \citet{psw+2016}. The ``swooshes'' of PSR B1859+07 which  lasts for about 20 to 120 periods, and happen gradually within several periods and somehow quasi-periodically for about every 150 rotations \citep{wor+2016, wyk+22}. In the normal state, PSR B1859+07 exhibits the bright and quiet modes \citep{rps+2021, wyk+22}, in which the central component is bright or weak respectively, which we term them as two sub-modes, i.e. central-Weak mode `cW' and central-Bright mode `cB'. The morphology, periodicity and possible physical origins of such a shift-like emission for PSR B0919+06 were also investigated, e.g. by \citet{hhp+2016}. These diverse features provide valuable insight into the physical conditions and emission processes in the pulsar magnetosphere.

There have been a number of interpretations on the ``swooshes'' phenomenon. Based on the symmetry of pulse profiles and cone-core beam model, \cite{rrw+2006} proposed that the ``swooshes'' are caused by partly illuminating of the emission cone, which might originate from the emission processes or `absorption' \citep{bkk+1981}. \citet{rps+2021} argued that the ``swooshes'' result from the cutting of different parts of flux tubes due to the shrinking and expanding of the magnetosphere. \citet{wor+2016} and \citet{gly+18} attributed the ``swooshes'' to the possible orbital dynamics in a binary system. These diverse interpretations need to be further verified, and the real origin of ``swooshes'' remains to be uncovered.

In this paper, we report the detection of a new emission mode of PSR B1859+07 by using the Five-hundred-meter Aperture Spherical radio Telescope (FAST), which has a clear implication for the origin of ``swooshes''. The paper is organized as follows. In section 2, FAST observations and data reduction are briefly introduced. The results for their morphology, polarization behaviours and subpulse drifting phenomenon are reported in section 3. Summary and discussion are given in Section 4.

\begin{figure*}
\centering
\includegraphics[height=0.83\textheight]{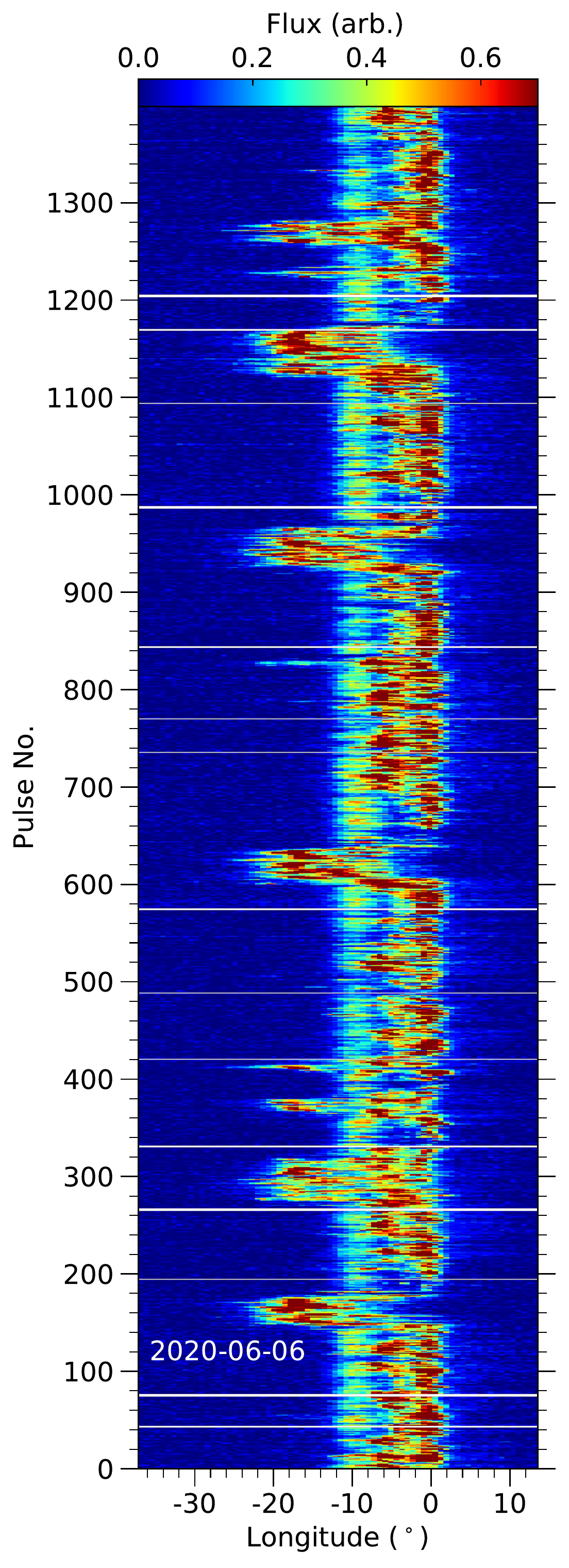}
\includegraphics[height=0.83\textheight]{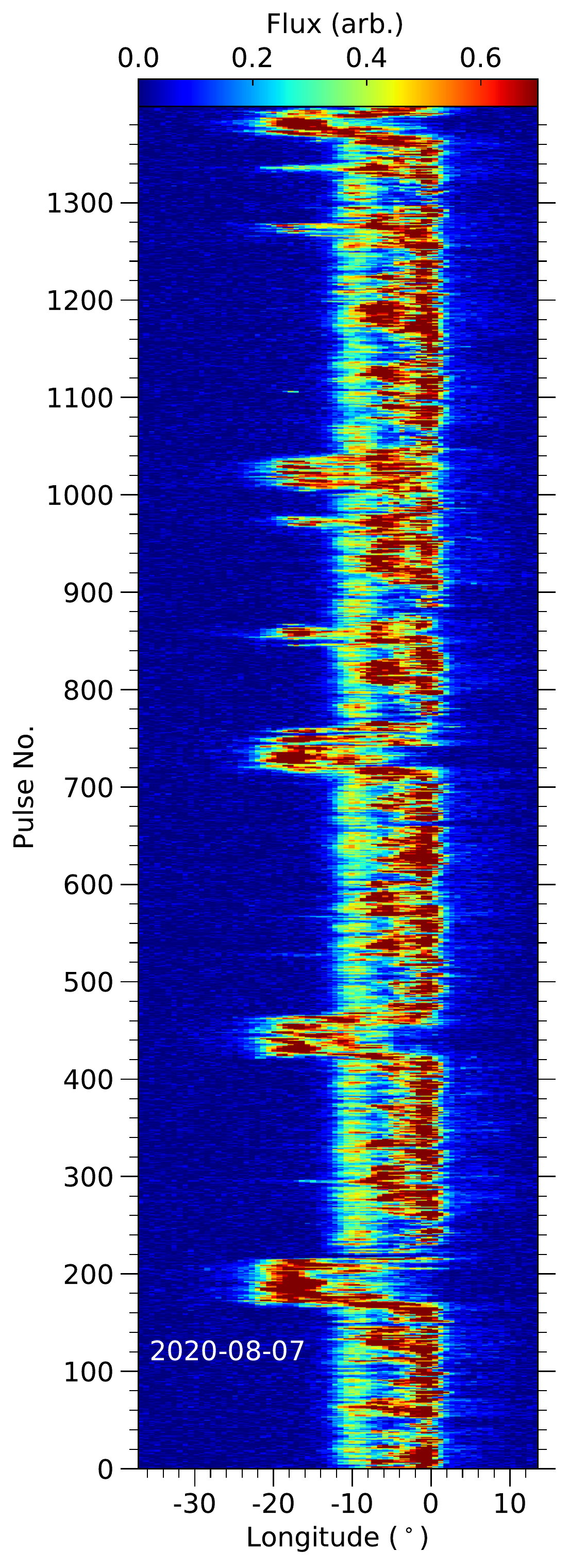}
\caption{Pulse sequences of PSR B1859+07 obtained from two FAST observations on June 06 and August 07, 2020 {in the GPPS survey}. Data with radio-frequency interference are eliminated as marked by white lines.
}
\label{figure:phase-time}
\end{figure*}

\begin{table}
    \caption[]{The period ranges for anomalous modes in two observations {of the FAST GPPS survey}.}
    \label{table:swooshlist}
\centering
    \begin{tabular}{llr}
    \hline
    \noalign{\smallskip}
    {\rm Period Range}      &   {\rm Type}  & {\rm Duration} \\ 
    {\rm (No.-No.)}  &   & {\rm (periods)}  \\ 
    \noalign{\smallskip}
    \hline
    \noalign{\smallskip}
    \multicolumn{3}{c}{20200606    } \\ \hline
                      153-172        &  A    &     19    \\
                      275-315        &  Af   &     40    \\
                      367-379        &  A    &     12    \\
                      409-415        &  Af   &     16    \\
                      494-495        &  Af   &     1     \\
                      606-630        &  A    &     24    \\
                      787-788        &  A    &     1     \\
                      824-828        &  Af   &     4     \\
                      930-955        &  A    &     25    \\
                      955-965        &  Af   &     10    \\
                      1120-1140      &  Af   &     20    \\
                      1140-1167      &  A    &     27    \\
                      1225-1229      &  A    &     4     \\
                      1256-1281      &  Af   &     35    \\
                      1332-1333      &  Af   &     1     \\
                      \hline
    \noalign{\smallskip}
   \multicolumn{3}{c}{    20200807} \\ \hline
                      170-216           &  A  &     46       \\
                      294-296           &  A  &     2        \\
                      425-455           &  A  &    30        \\
                      455-465           &  Af &    10        \\
                      714-720           &  Af &     6        \\
                      720-742           &  A  &    22        \\
                      742-760           &  Af &    18        \\
                      845-867           &  Af &    22        \\
                      968-977           &  Af &     9        \\
                      1006-1038         &  Af &    32        \\
                      1104-1106         &  Af &     2        \\
                      1268-1278         &  Af &    10        \\
                      1332-1338         &  Af &     6        \\
                      1367-1388         &  A  &    21        \\
                      1388-1395         &  Af &     7        \\ 
    \noalign{\smallskip}
    \hline
    \end{tabular}
\end{table}

\section{Observations and data reduction}
\label{data}

We made two observations of PSR B1859+07 on June 6th, 2020 (20200606) and August 7th, 2020 (20200807) with the 19-beam receiver of the FAST \citep{jth+20}, during the verification  observations for the FAST Galactic Plane Pulsar Snapshot (GPPS) survey \citep{HWW+2021}. Each tracking observation lasts for 15 minutes. The data was recorded with full polarization, which has a central frequency of 1.25GHz and a bandwidth of 500MHz. The data for 2048 frequency channels are stored with a time resolution of 49.152 microsecond. Details of observations can be found in \citet{HWW+2021}. Before each observation session, the two minutes data are recorded from the receiver with  calibration signals of 1~K On-Off noise injected every 2~s, which are used for the correction of the band-pass of the receiver and also the calibration of receiver polarization performance. 

The special targeted observation to PSR B1859+07 on 20191203 and 20201122 \citep{wyk+22} are also included in this work.

Offline data processing is as follows. We first dedispersed the data and formed single pulse sequences with the ephemeride by using DSPSR \citep{vb11}. Radio frequency interference was then exercised for frequency channels and time by using PSRCHIEVE \citep{hvm04}. The pulses were finally calibrated in polarization with noise diodes.

\begin{figure*}
\centering
\includegraphics[width=0.3\linewidth]{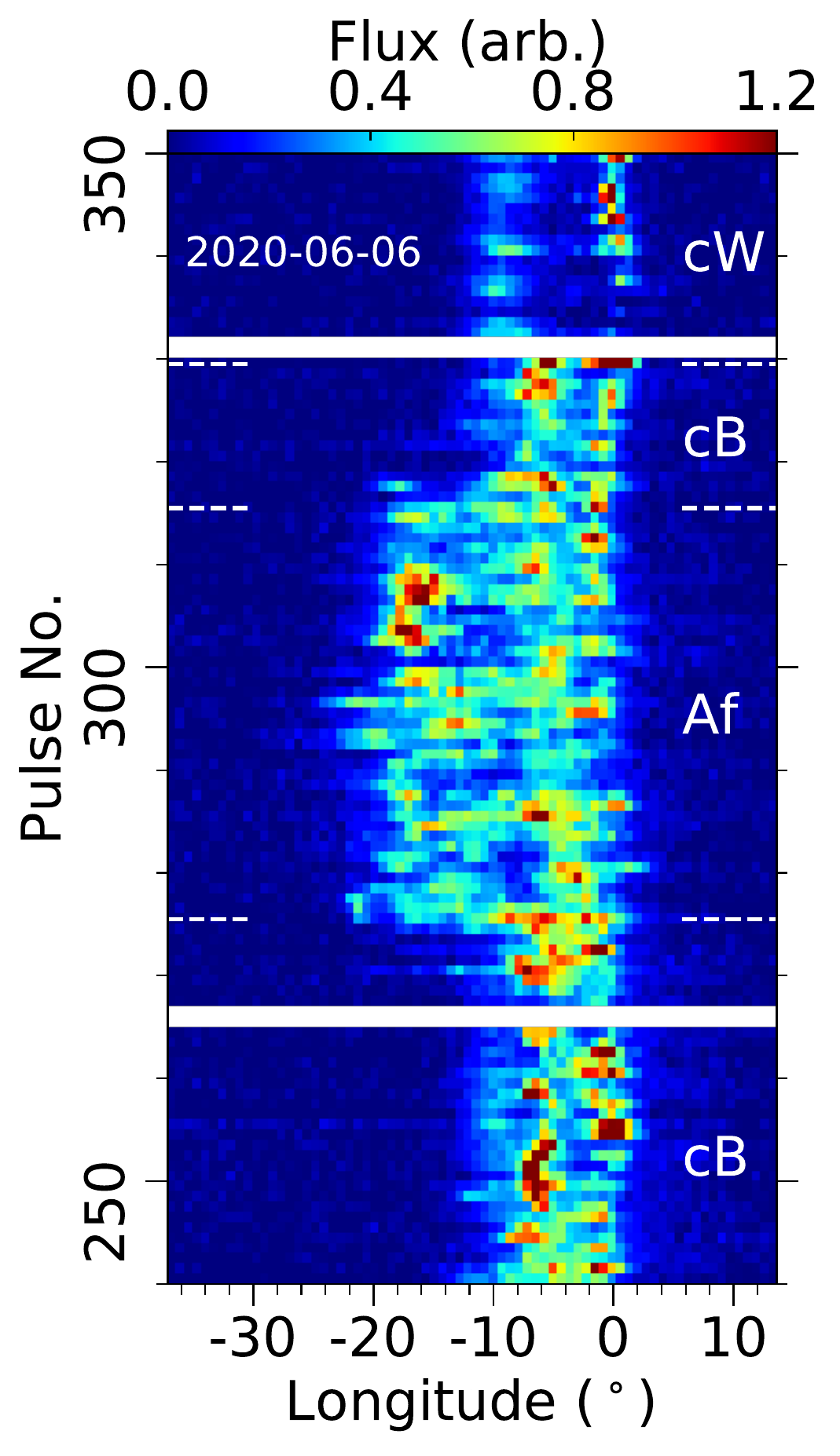}
\includegraphics[width=0.3\linewidth]{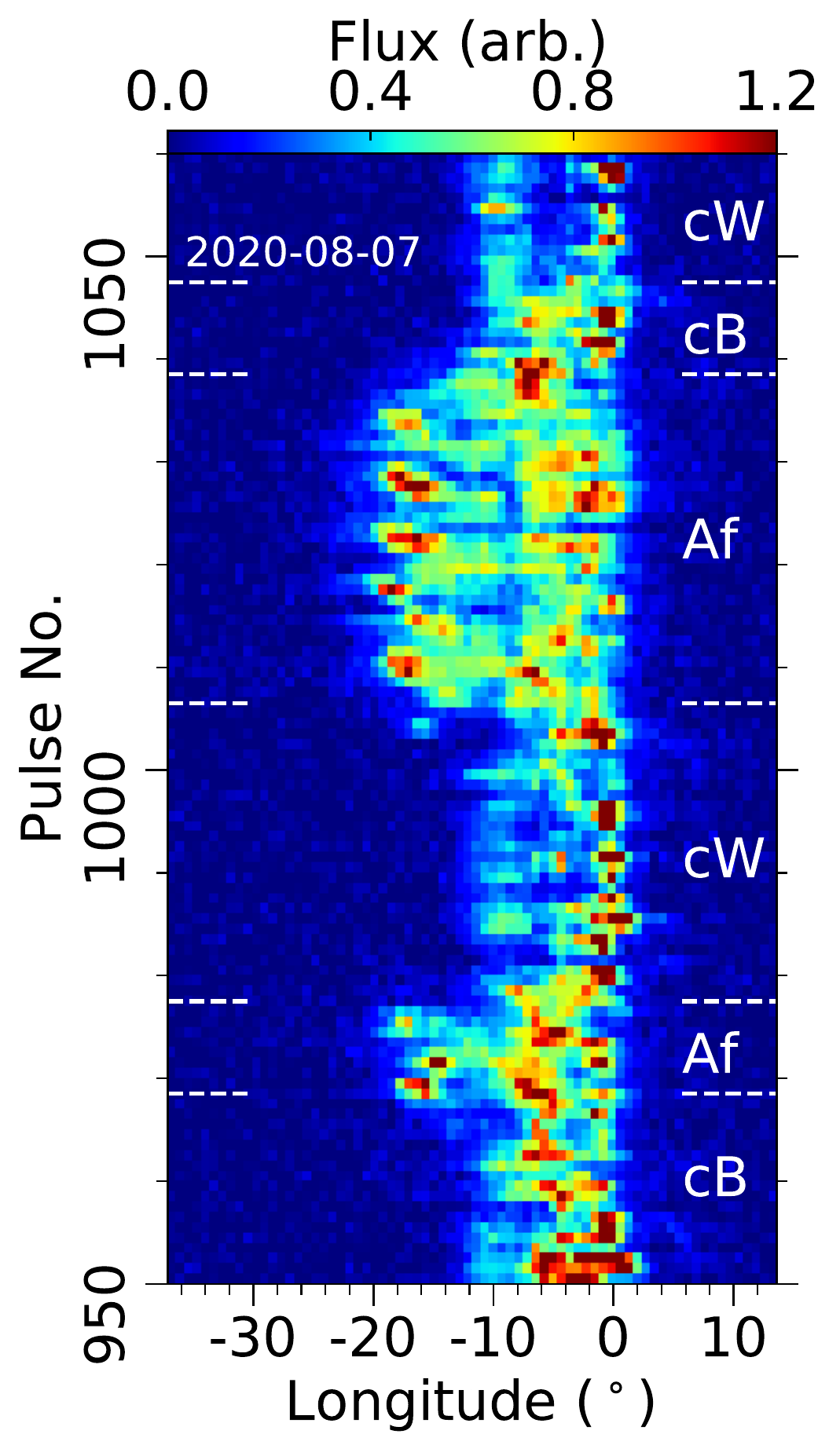}
\includegraphics[width=0.3\linewidth]{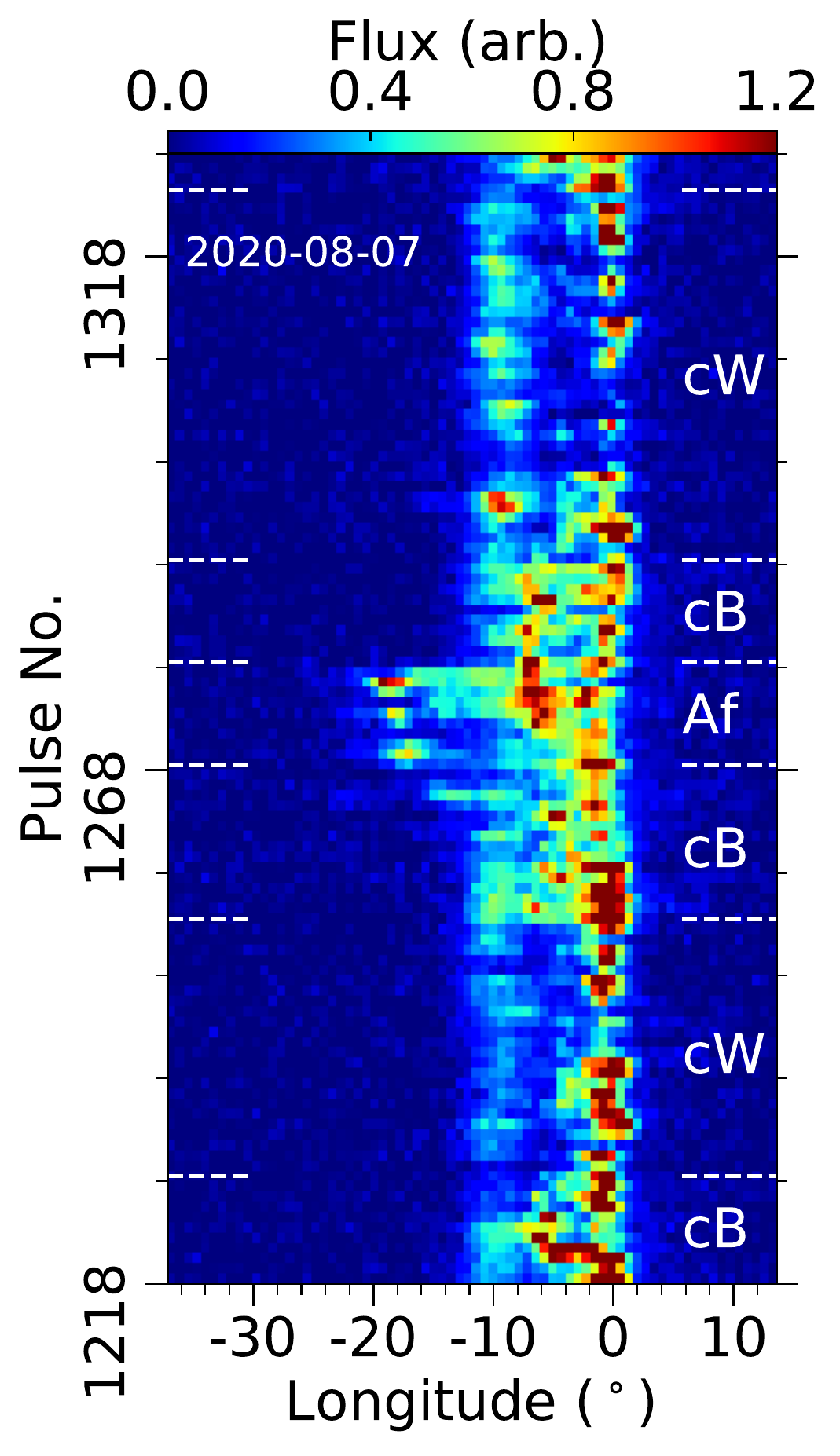}
\caption{Examples for the Af mode, which show not only the new leading component in the A mode but also have the slightly squished components of the cB and cW modes. The different emission modes are marked for different segments of pulse sequences. The data with radio-frequency interference are eliminated as marked by white lines.
}
\label{swooshType}
\end{figure*}

\section{Results}

The pulse sequences are obtained from the two observations in the GPPS survey on June 06 and August 07, 2020, as shown in Figure~\ref{figure:phase-time}. The pulse sequences from special targeted FAST observations PSR B1859+07 on 20191203 and 20201122 have been shown in \citep{wyk+22}.
Obviously different emission modes switch frequently with various duration. As previously observed \citep{rrw+2006,wor+2016,rps+2021}, the normal emission mode has three prominent components in the averaged profile, and the trailing one is the brightest. According to the brightness of the central component, the normal emission mode can be further classified into two sub-modes, one with a bright central component as the {\it cB mode }corresponding to the ``B mode'' in \citet{wyk+22} and \cite{rps+2021}, and the other with a weak central component as the {\it cW mode} corresponding to the ``Q mode'' in \citet[i.e.][]{wyk+22} or the ``A mode'' in \cite{rps+2021}. The profile and polarization differences of these two modes have been studied in great details by \citet{wyk+22} and \cite{rps+2021}. What we are concerned about in this paper is the ``swooshes" which exhibits a new bright component in the advanced phase range of $-12^{\circ}$ to $-28^{\circ}$. The period ranges for the  anomalous mode are listed in Table~\ref{table:swooshlist}.  

\begin{figure*}
\centering
\includegraphics[width=0.37\linewidth]{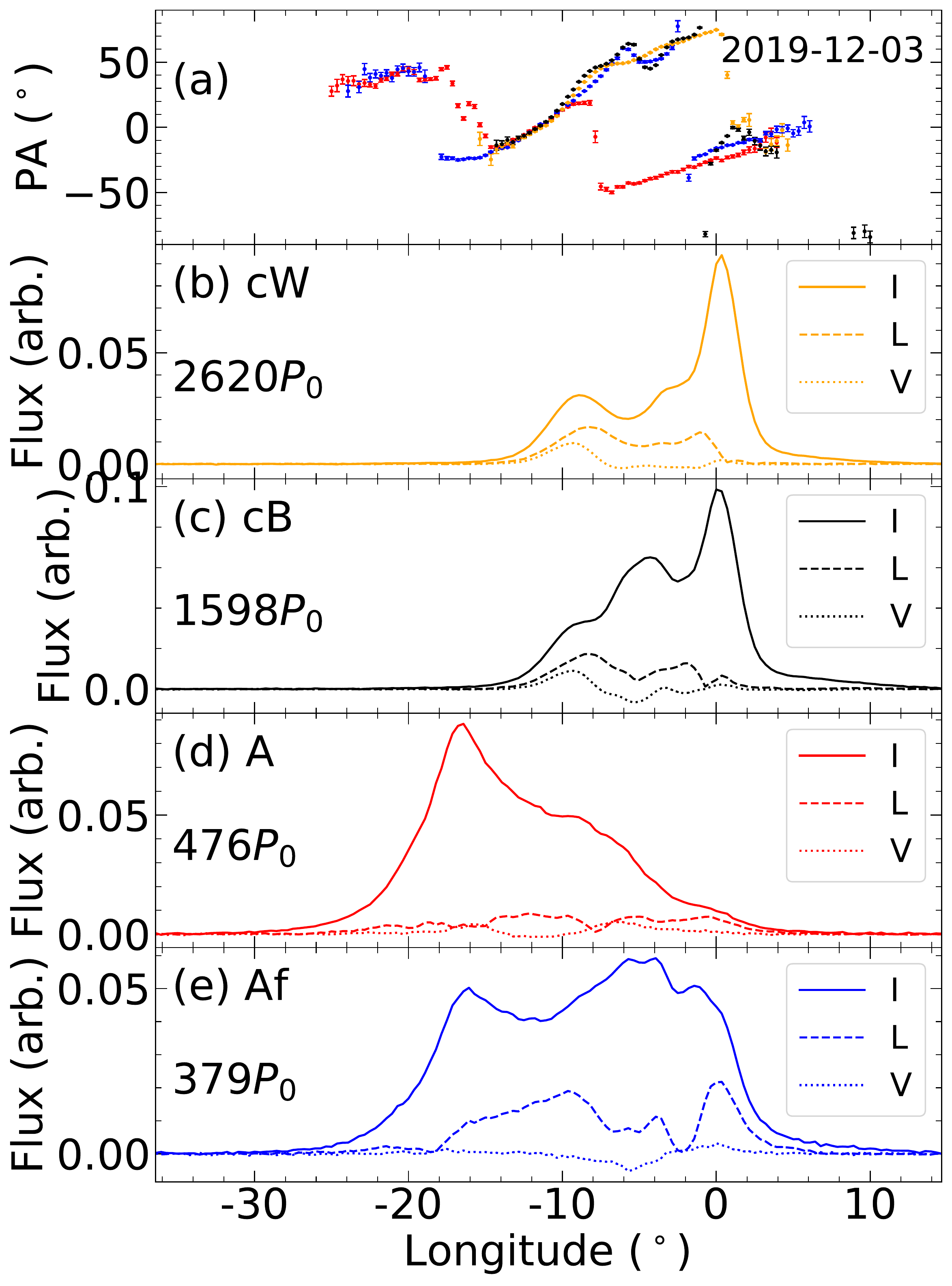} 
\includegraphics[width=0.37\linewidth]{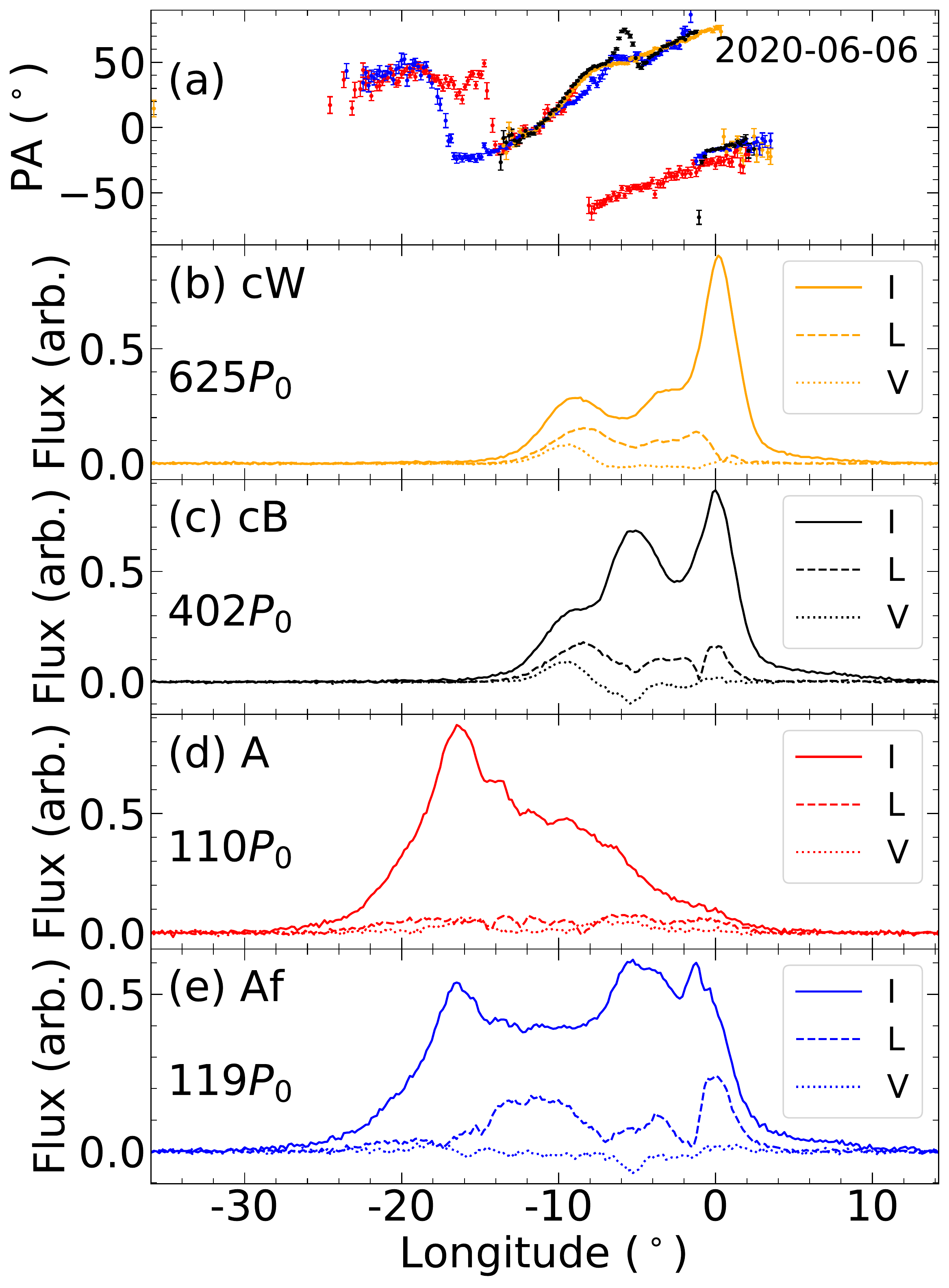} 
\includegraphics[width=0.37\linewidth]{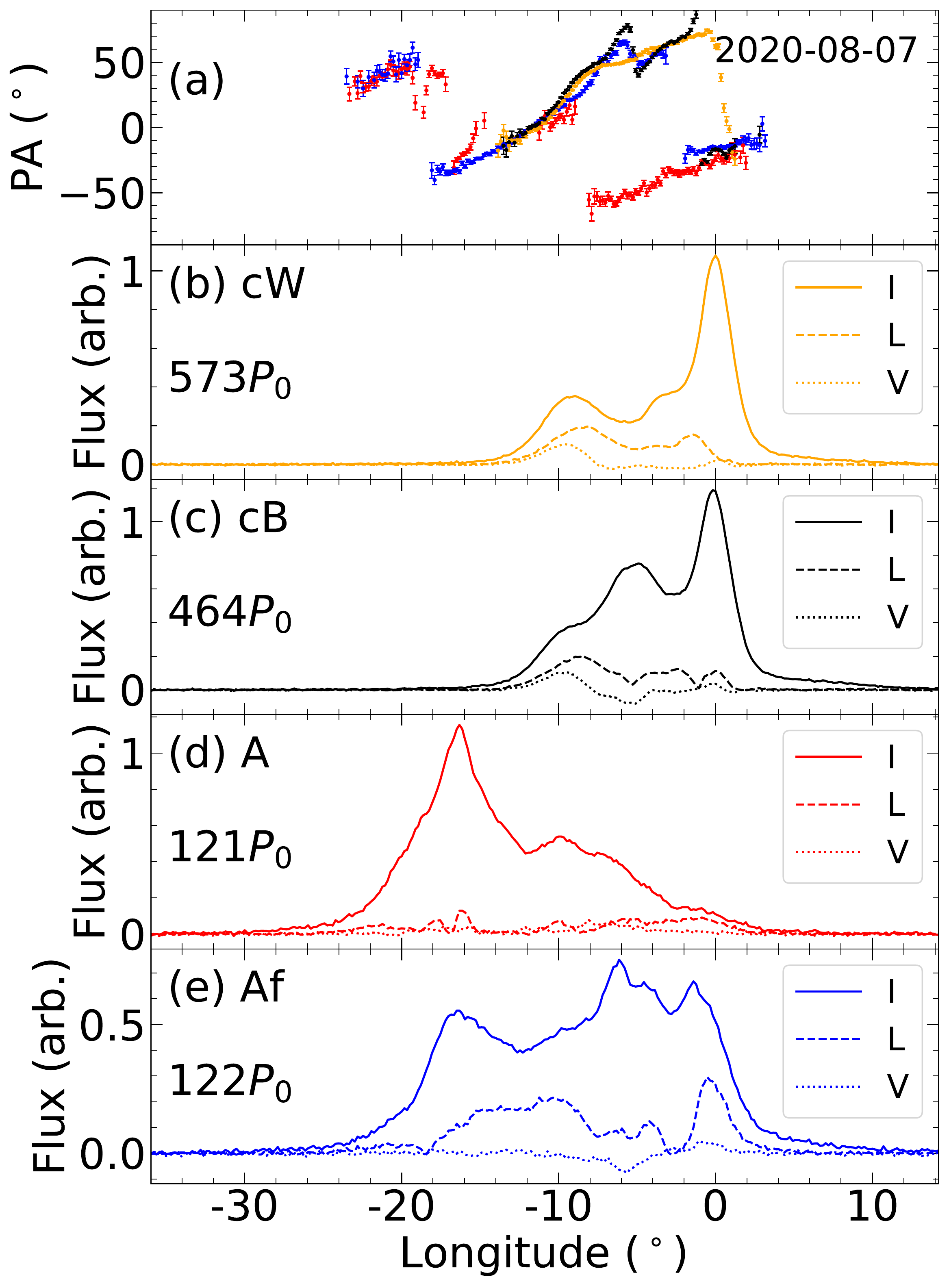}
\includegraphics[width=0.37\linewidth]{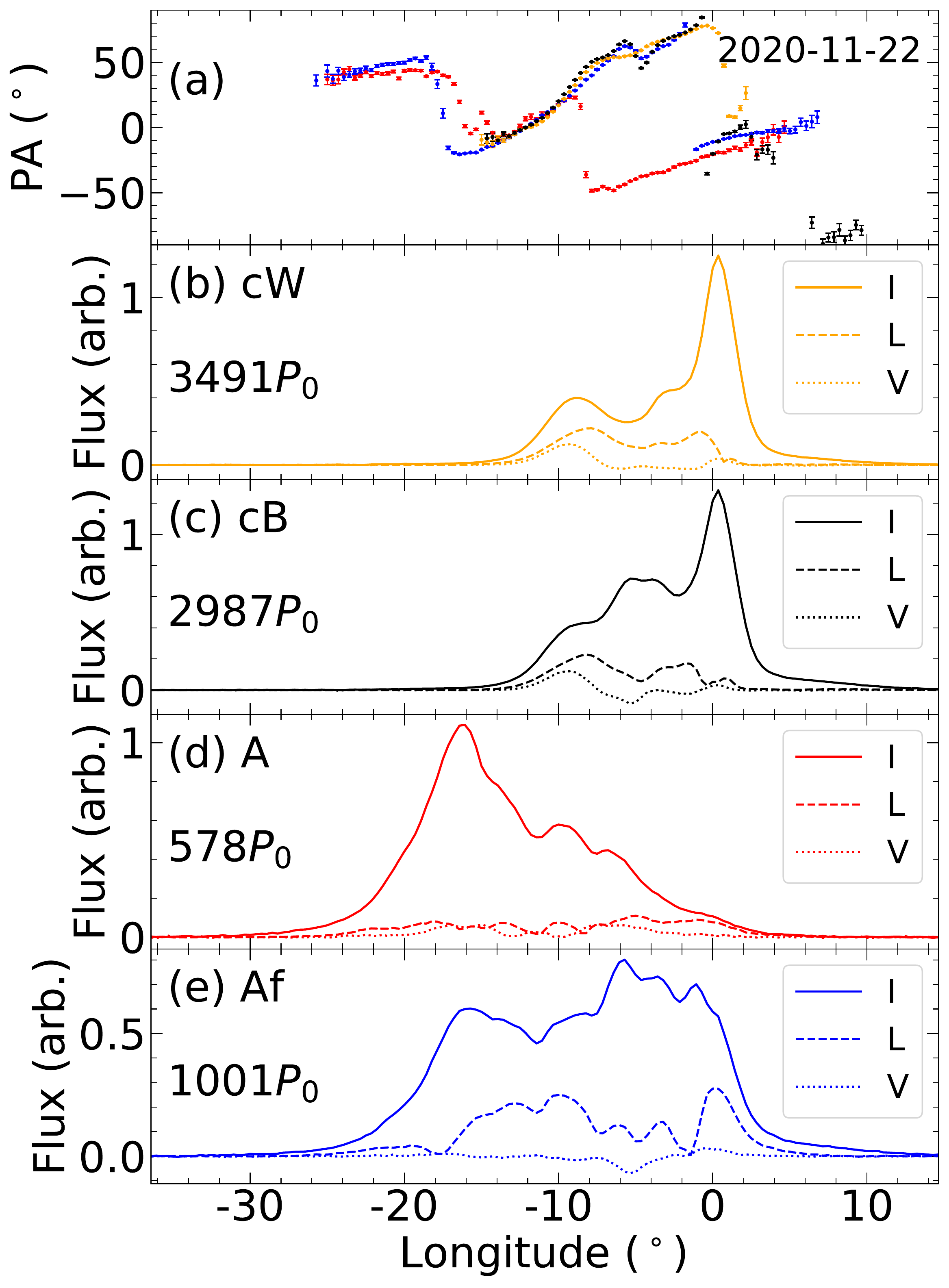}
\caption{The mean pulse profiles of the emission modes: cW (the orange), cB (the black), A (the red) and Af (the blue), from 4 FAST observations, with the number of pulsar periods marked in each subpanel. In general, profiles for the normal mode are consistent with each other in the  four observations. The slight difference for the Af mode profiles are caused by insufficient number of periods for the average process.  The PA curves in the Af mode are in the orthogonal mode of those in the A mode in the phase range of [-8.5$^\circ$,-1.5$^\circ$]. The position angle  data are plotted only when linear polarization intensity exceeds 5 times the standard deviation of its off-pulse emission.
}
\label{figure:poln3mode}
\end{figure*}

\begin{figure*}
\centering
\includegraphics[width=0.42\linewidth]{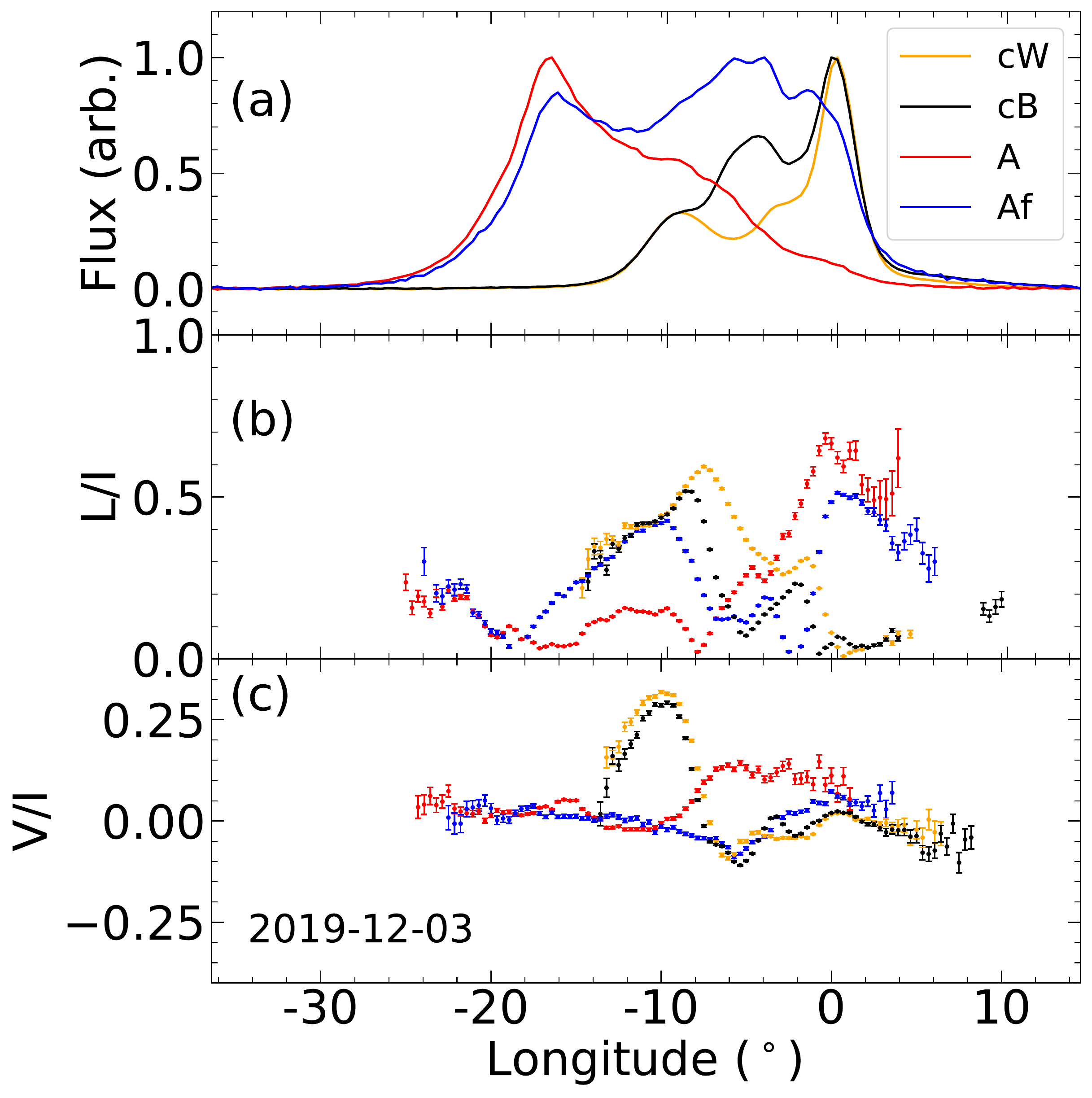} 
\includegraphics[width=0.42\linewidth]{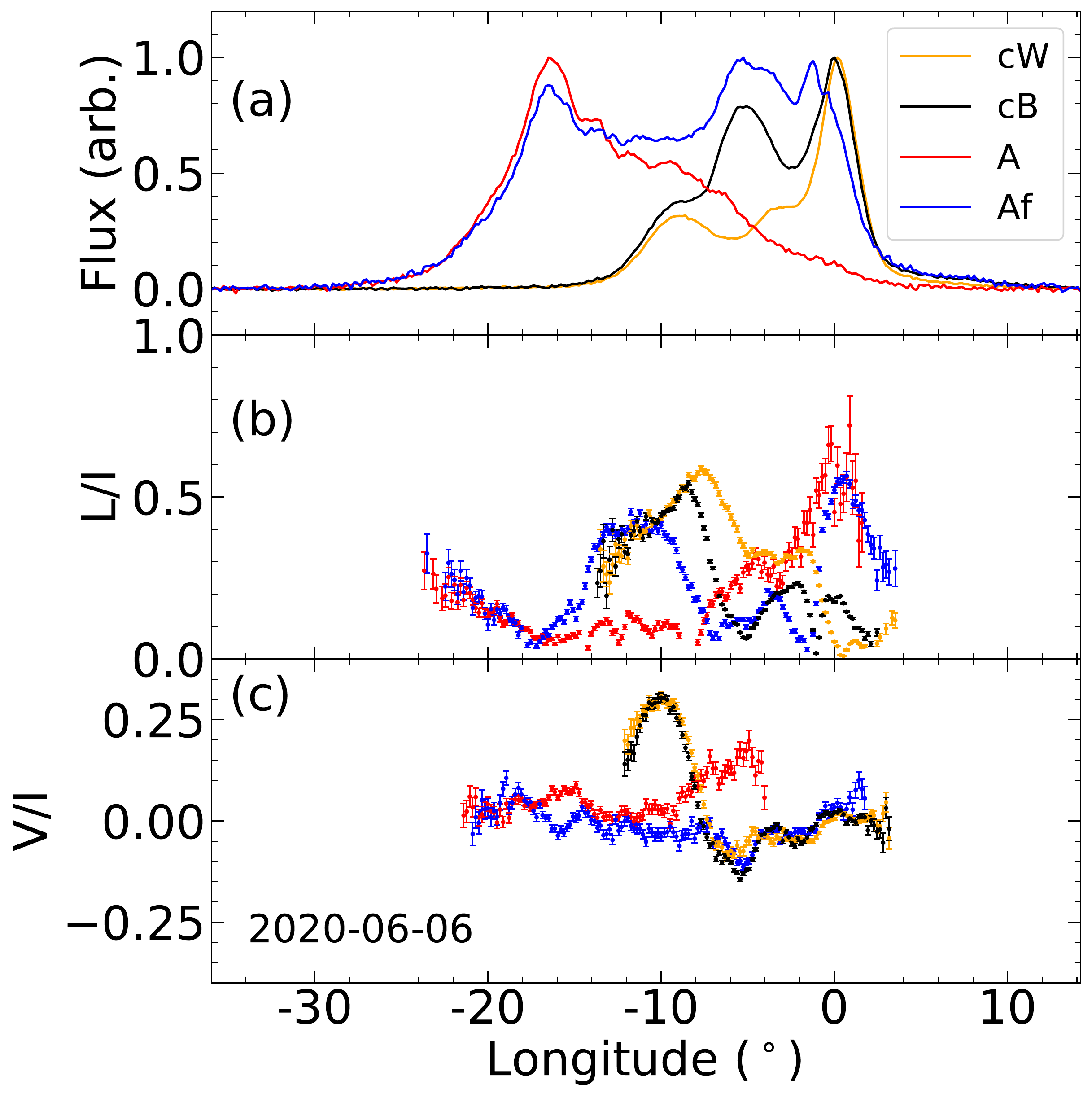} 
\includegraphics[width=0.42\linewidth]{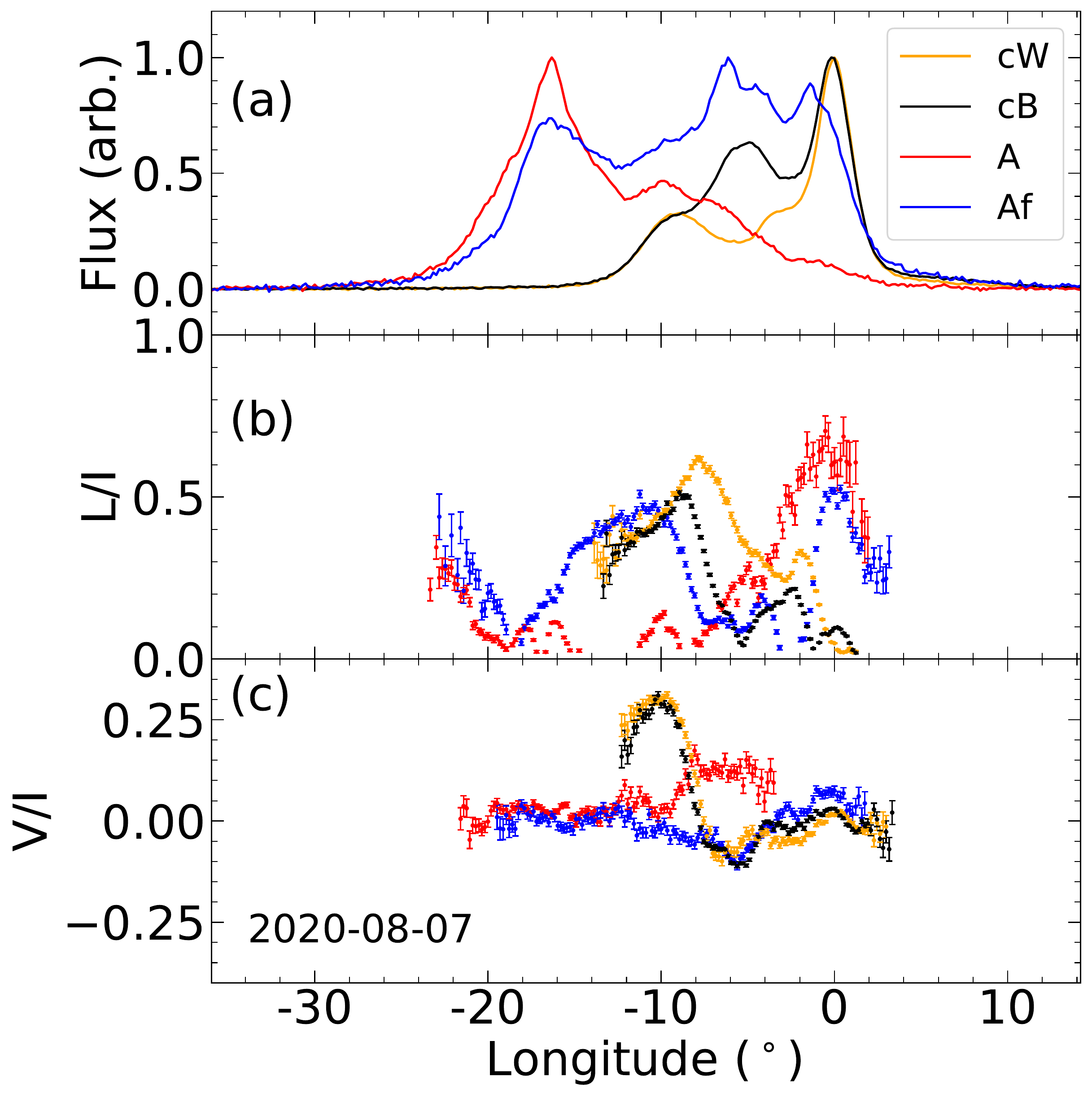}
\includegraphics[width=0.42\linewidth]{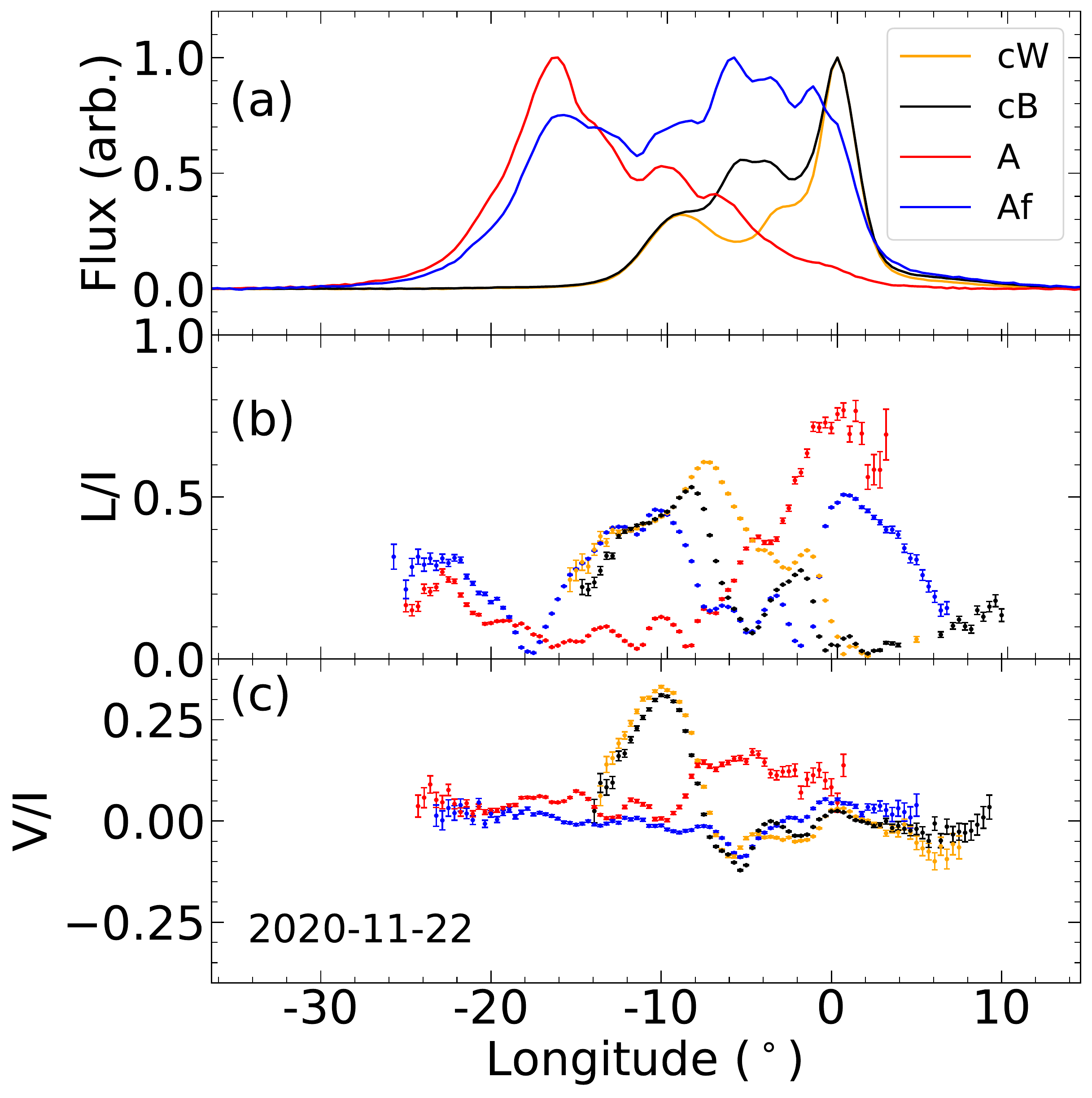}
\caption{Comparison of mean pulse polarization profiles for the four modes (cW, cB, A and Af) obtained from 4 FAST observations. The total intensity, fractional linear and circular polarization are plotted in the different subpanels from top to bottom. The fractional linear polarization data are plotted only when linear polarization intensity exceeds 5 times the standard deviation of its off-pulse emission, so do the fractional circular polarization. 
}
\label{figure:comp3modepoln}
\end{figure*}

\subsection{The anomalous-filled mode: ``Af mode''}

As shown in Figure~\ref{figure:phase-time}, there are two types of  ``swooshes" revealed by our sensitive FAST observations. 
One is the conventional ``swooshes'' in which a new bright component emerges in the advanced phase range while the bright trailing component seen in the normal mode completely disappears as if the emission components are shifted to an earlier phase. We now label it as the {\it mode A}. 
The other one is not such ``swooshes'' but appears as a new emission mode, dubbed as the {\it Af mode}, which hosts all emission components seen from the normal mode and anomalous mode, though the leading component (almost the same one in the anomalous ``A mode'') and the trailing component (almost the same one in the normal mode) are shrinked in the longitude phases. Examples for such an anomalous-filled mode, i.e. ``the Af mode'', are shown {in} Figure~\ref{swooshType}. 

We identify 30 anomalous emission events from our two FAST  observations in the GPPS survey, lasting for 1 to 46 or 40 rotations. including 12 events for the ``A mode'' and 18 events for the ``Af mode'', as listed in Table~\ref{table:swooshlist}. In fact, such an anomalous-filled mode has been noticed and shown in the FAST data { on 20201122 published by} \citet{wyk+22}, e.g. the pulses around period No. 450, 1270, 1580, 1990, 2140, 2700, 3170, 3510 and 4570 in their Fig.2, and the upper half of their Fig.6 gives the best demonstration, but this mode was not well investigated especially on the polarization properties. We also checked FAST observation results on 20191203, and found many segments for the Af mode.

\subsection{Polarization profiles for different modes}

Polarization profiles of pulsars provide insights into the emission geometry in pulsar magnetosphere. Comparing polarization profiles for different modes, especially the polarization angle curves, can aid in distinguishing between ``swooshes" in different longitudinal phases and different modes of pulsar emission.

By averaging the pulses in the different modes, we get the mean pulse polarization profiles for the two FAST observations in Figure~\ref{figure:poln3mode}.The polarization profiles for the cW mode and the cB mode have been compared by \citet{wyk+22} already. We here compare the polarization profiles for two normal modes (``cW'' and ``cB") with the two anomalous modes (``A" and `Af"). The profiles for all modes are consistent with each other in the two observations. The slight profile difference between the 4 observations for the A mode and the Af mode reflects the unstable profiles caused by insufficient numbers of periods for the average process. Very striking is the PA curves for the Af mode, which are in the orthogonal mode of that for the A mode in the phase range between -8.5$^\circ$ and -1.5$^\circ$. In general, the PA curves of the Af mode follow the curves of the cB and cW modes except the phase range around 0$^\circ$.

One may compare the polarization profiles by plotting them together as in Figure~\ref{figure:comp3modepoln}. The results for the fractional linear polarization and the fractional circular polarization are very consistent with each other in the two days. 
The fractional linear polarization of the ``A mode'' emission differs significantly from that of the normal ones and the ``Af mode'', with an extremely lower percentage in the phase range of $-15^\circ$ to $-7^\circ$.
For the emission of ``Af" mode, the fractional linear polarization is much higher at the leading and trailing part of the profiles, but it is relatively lower within the phase range of  $-8^\circ$ to $-5^\circ$. 
The very different fractional circular polarization is seen in the phase range of  $-14^\circ$ to $-5^\circ$. The normal modes have a sense-reversal at around -8$^\circ$, while the ``mode A'' has a circular polarization peak at the phase of $-5^\circ$ while ``mode Af'' has a dip. In the other longitude ranges the circular polarization remains the same, and the circular polarization of the Af mode obviously follows the curve for the leading components of the A mode and the trailing components of the cW and cB modes.

All these polarization features suggest that the Af mode is a good ``combination'' of the ``swooshes" of emission and two normal emission modes, indicating that the A mode is not the swooshed normal emission mode.

\begin{figure}
\centering
\includegraphics[width=0.40\textwidth]{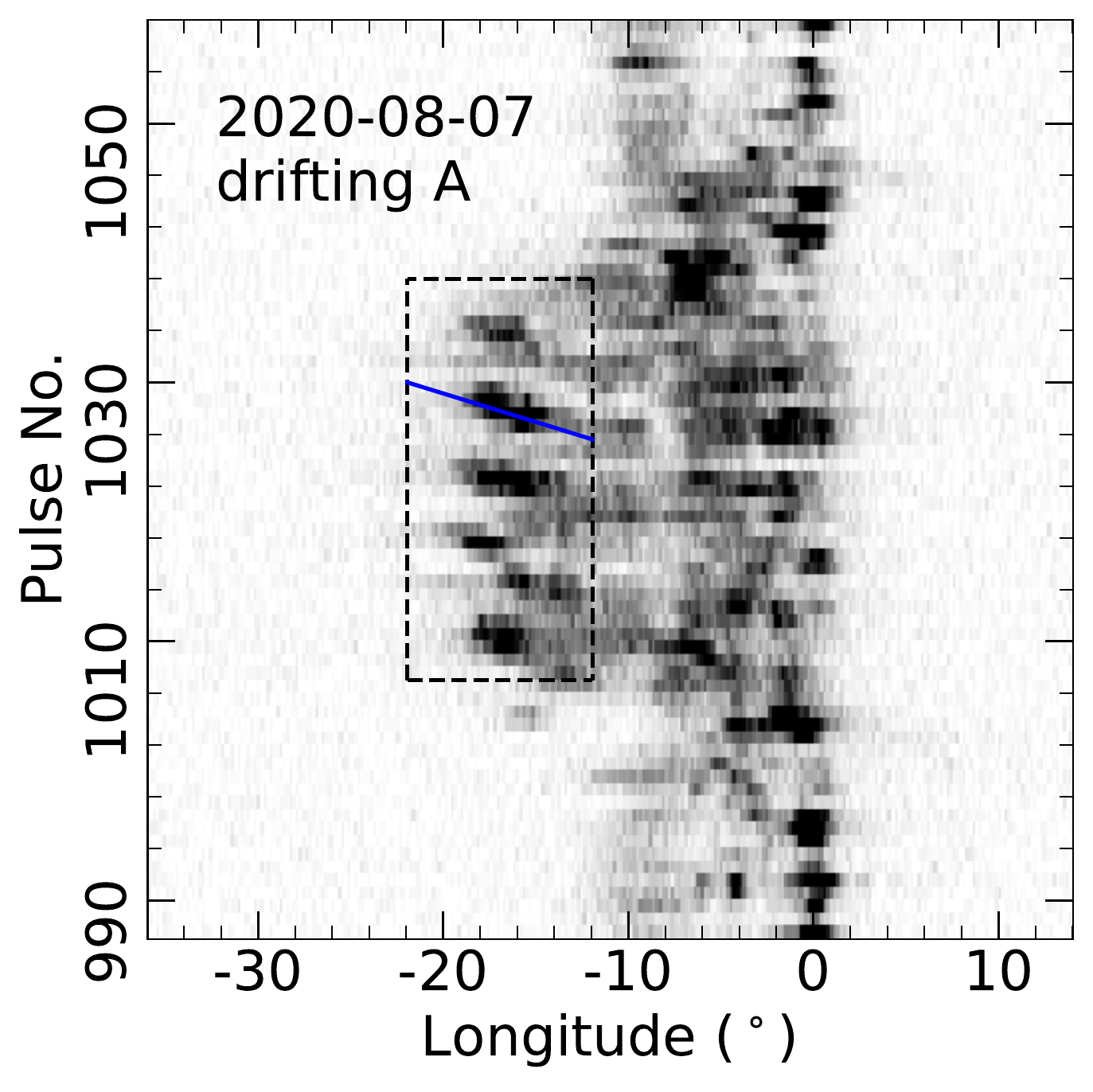}
\caption{Sub-pulse drifting of PSR B1859+07 
{in the "Af" mode.} The blue line indicates the drifting direction. }
\label{figure:drifting}
\end{figure}

\subsection{Subpulse drifting in the new leading component of  the Af mode}

 One, only one, subpulse drifting event is identified from the new leading component in the Af mode, as shown in Figure~\ref{figure:drifting}. 

We fit the emission peaks {in the phase range of $-20^\circ$ to $-12^\circ$} in these periods of the ``Af mode'', and we get the drifting rate as being $P_2/P_3 = -2.3^\circ$ per period. The period gap for a sub-pulse appearing at the same longitude, that is $P_3$, is measured to be 6 periods. The longitude spacing between two adjacent sub-pulses, $P_2$, derived from the cross-points of two drifting bands in a given period, is found to be $-13.8^\circ$.

\section{Summary and discussion}

In summary, based on the new FAST observations of PSR B1859+07, we find following features for different modes: 

(1) A new emission mode: We identify the ``Af mode'' which is different from the previously known ``swooshes'' and has all components from the normal emission mode and anomalous ``swooshes'', which means that the emission components of the ``Af mode'' is not the swooshed components of normal emission, so that all such related theoretical interpretation involving ``swoosh'' should be abandoned.  

(2) The duration for the anomalous emission mode: That should vary from a single rotation {period} to a few tens, as seen in Table 1.

(3) Emission phase ranges: Because the newly identified Af mode shows a much wide phase range for the emission, with the PA curves connecting PAs of the leading component in the anomalous mode and the trailing component in the normal emission mode, the normal mode should then be only a part of illuminated conal emission, as if the partial cone discussed by \citet{lm+1988,rrw+2006}. 

(4) Orthogonal modes: The polarization angles for the ``Af mode'' are coincident with the orthogonal mode of the ``A mode'' in some phase range, implying that they experience different propagation effects. 

(5) Circular polarization: The circular polarization of normal mode emission changes the sense around the phase of $-8^{\circ}$, supporting the explanation of the normal emission profile as the partially illuminated conal emission \citet{rps+2021}. 

(6) Subpulse drifting: We get the first detection of drifting subpulses for this pulsar in the leading component of the Af mode, and derive the drifting parameters from one drifting session. More data are desired for further investigation.

There have been a number of interpretations for the anomalous mode of PSR B0919+06 and B1859+07. The models rested on aberration effects and binary interactions have already been excluded by \citet{rrw+2006} and \citet{wyk+22}. The polarization features of the newly identified mode of PSR B1859+07 favour the simple model in which the anomalous emission events are not caused by the so-called phase shift, but just the intrinsic radiation of different parts of pulsar emission beam. The orthogonal mode between the different modes suggests possible changes on the energy and density distributions of relativistic particles for propagation effects in the entire pulsar magnetosphere\citep{wwh14}. Since partial cones have been observed for many pulsars already \citep{lm+1988}, which depends on sight line geometry, the relativistic particles and various emission processes and propagation effects, caused probably by dynamical sparking in pulsar polar cap. The subpulse drifting occasionally observed and the sense-change of circular polarization are also key observational facts for understanding the anomalous mode of these pulsars.

\section*{Acknowledgements}

FAST is a Chinese national mega-science facility built and operated by
the National Astronomical Observatories, Chinese Academy of
Sciences.
P. F. Wang is supported by the National Key R\&D Program of China
(No. 2021YFA1600401 and 2021YFA1600400), National Natural Science
Foundation of China (No. 11873058 and 12133004).
J. L. Han is supported by the National Natural Science
Foundation of China (No. 11988101 and 11833009).

\section*{DATA AVAILABILITY}

Original FAST observation data are accessible under the FAST data open policy, i.e. full available one year after observations. All processed data as plotted in this paper can be obtained from the authors with a kind request.





\bibliographystyle{raa}
\bibliography{citation}





\label{lastpage}

\end{document}